\documentclass[conference]{IEEEtran}
\IEEEoverridecommandlockouts
\usepackage{cite}
\usepackage{amsmath,amssymb,amsfonts}
\usepackage{algorithmic}
\usepackage{graphicx}
\usepackage{textcomp}
\usepackage[table,xcdraw]{xcolor}
\usepackage{enumitem}
\usepackage{tikz}
\usepackage{url}
\usepackage{booktabs}
\newcommand*\circled[1]{\tikz[baseline=(char.base)]{%
            \node[shape=circle,draw,inner sep=1pt] (char) {#1};}}

\def\BibTeX{{\rm B\kern-.05em{\sc i\kern-.025em b}\kern-.08em
    T\kern-.1667em\lower.7ex\hbox{E}\kern-.125emX}}
\begin{document}

\newcommand\changed[1]{\textcolor{blue}{#1}}
\newcommand\todo[1]{\textcolor{red}{#1}}
\newcommand\revised[1]{\textcolor{blue}{#1}}

\renewcommand\changed[1]{#1}
\renewcommand\todo[1]{#1}
\renewcommand\revised[1]{#1}

\title{CAGRA: Highly Parallel Graph Construction and Approximate Nearest Neighbor Search for GPUs
}

\makeatletter
\newcommand{\linebreakand}{%
  \end{@IEEEauthorhalign}
  \hfill\mbox{}\par
  \mbox{}\hfill\begin{@IEEEauthorhalign}
}
\makeatother

\author{\IEEEauthorblockN{1\textsuperscript{st} Hiroyuki Ootomo}
\IEEEauthorblockA{\textit{NVIDIA} \\
Tokyo, Japan \\
ORCID: 0000-0002-9522-3789}
\and
\IEEEauthorblockN{2\textsuperscript{nd} Akira Naruse}
\IEEEauthorblockA{\textit{NVIDIA} \\
Tokyo, Japan \\
ORCID: 0000-0002-3140-0854}
\and
\IEEEauthorblockN{3\textsuperscript{rd} Corey Nolet}
\IEEEauthorblockA{\textit{NVIDIA} \\
Maryland, USA \\
ORCID: 0000-0002-2117-7636
}
\linebreakand
\IEEEauthorblockN{4\textsuperscript{th} Ray Wang}
\IEEEauthorblockA{\textit{NVIDIA} \\
Shanghai, China \\
ORCID: 0000-0001-8982-0571
}
\and
\IEEEauthorblockN{5\textsuperscript{th} Tamas Feher}
\IEEEauthorblockA{\textit{NVIDIA} \\
Munich, Germany \\
ORCID: 0000-0003-2095-4349}
\and
\IEEEauthorblockN{6\textsuperscript{th} Yong Wang}
\IEEEauthorblockA{\textit{NVIDIA} \\
Shanghai, China \\
ORCID: 0009-0005-0906-8778}
}

\maketitle

\begin{abstract}
Approximate Nearest Neighbor Search (ANNS) plays a critical role in various disciplines spanning data mining and artificial intelligence, from information retrieval and computer vision to natural language processing and recommender systems.
Data volumes have soared in recent years and the computational cost of an exhaustive exact nearest neighbor search is often prohibitive, necessitating the adoption of approximate techniques.
The balanced performance and recall of graph-based approaches have more recently garnered significant attention in ANNS algorithms, however, only a few studies have explored harnessing the power of GPUs and multi-core processors despite the widespread use of massively parallel and general-purpose computing.
To bridge this gap, we introduce a novel parallel computing hardware-based proximity graph and search algorithm.
By leveraging the high-performance capabilities of modern hardware, our approach achieves remarkable efficiency gains.
In particular, our method surpasses existing CPU and GPU-based methods in constructing the proximity graph, demonstrating higher throughput in both large- and small-batch searches while maintaining compatible accuracy.
In graph construction time, our method, CAGRA, is $2.2 \textendash{} 27\times$ faster than HNSW, which is one of the CPU SOTA implementations.
In large-batch query throughput in the 90\% to 95\% recall range, our method is $33\textendash{} 77\times$ faster than HNSW, and is $3.8\textendash{} 8.8\times$ faster than the SOTA implementations for GPU.
For a single query, our method is $3.4\textendash{} 53\times$ faster than HNSW at 95\% recall.
\end{abstract}


\section{Introduction}
\label{sec:introduction}
%
The importance of Approximate Nearest Neighbor Search (ANNS) has grown significantly with the increasing volume of data we encounter. ANNS is particularly relevant in solving the $k$ Nearest Neighbor Search ($k$-NNS) problem, where we seek to find the $k$ vectors closest to a given query vector, typically using a distance like the L2 norm, from a dataset of vectors.
The simplest exact solution for $k$-NNS involves exhaustively calculating the distance between the query vector and all vectors in the dataset, then outputting the $k$ vectors of the smallest distances (top-$k$) as results.
However, this approach becomes infeasible for large datasets due to the sheer number of similarity computations required, making scalability an issue. In many practical applications, exact results are not always necessary and ANNS can strike a balance between throughput and accuracy, reducing the computational burden and enabling the scaling to large datasets.
The impact and use-cases for ANNS are widespread and include several disciplines spanning data mining and artificial intelligence, such as language models in natural language processing \cite{khandelwal_generalization_2020,xu_why_2023}, computer vision \cite{kanji_mining_2014,banerjee_view_2019}, information retrieval \cite{liu_clustering_2007}, recommender systems, and advertising \cite{li_practice_2023,chen_approximate_2022}.
ANNS also forms the core of many important classes of data science and machine learning algorithms such as clustering\cite{nolet_cuslink_2023}, classification \cite{nolet_gpu_2022}, manifold learning and dimensionality reduction \cite{nolet_bringing_2021}.
Various different categories of algorithms for ANNS have been proposed and are well-studied, including hashing-based \cite{datar_locality-sensitive_2004}, tree-based \cite{silpa-anan_optimised_2008}, quantization-based \cite{jegou_product_2011}, and graph-based methods.

%
The graph-based method for ANNS relies on a \textit{proximity graph}, or a graph that represents the similarity relationships among data points within a dataset.
Graph-based methods involve two primary steps: constructing a proximity graph from a dataset and traversing it to find the $k$ closest nodes to the input query.
The question of determining the optimal proximity graph structure is not easily answered theoretically.
For instance, the graph quality as a $k$-nearest neighbor graph does not necessarily guarantee higher accuracy \cite{wang_comprehensive_2021}.
Therefore, researchers have focused on optimizing the graph's efficiency and structure heuristically and empirically.
One notable approach is the NSW (Navigable Small World) graph proposed by Malkov et al. \cite{malkov_approximate_2014}, which leverages the Small World phenomenon \cite{travers_experimental_1969} to enhance the search performance of the proximity graph. Building upon this idea, HNSW (Hierarchical Navigable Small World) \cite{malkov_efficient_2018} graphs address issues present in NSW, where a few nodes have a large degree, hindering the high-performance search. 
HNSW addresses this problem by introducing a hierarchical graph structure and setting an upper bound on the maximum degree.
Another approach is NSG (Navigating Spreading-out Graph), proposed by Fu et al. \cite{fu_fast_2019}, which approximates a Monotonic Relative Neighborhood Graph (MRNG) structure to help guarantee low search complexity.
These are just a few examples of graph structures used in ANNS, and numerous other graphs have been well-studied in the field \cite{wang_comprehensive_2021}.

%
When it comes to implementing graph-based ANNS methods, few studies have introduced high-performance implementations optimized for data-center servers capable of harnessing the massive parallelism offered by GPUs.
\revised{The memory bandwidth to load the dataset vectors can be a bottleneck of the search throughput, not only in graph-based, but also in other ANNS implementations. Since a GPU, typically has high memory bandwidth, it is potentially suitable for ANNS.}
One notable implementation is SONG \cite{zhao_song_2020}, which stands as the first graph search implementation on GPUs, using various optimization techniques.
These techniques include employing the open addressing hash table and performing multi-query searches within a warp.
\revised{With these optimizations, SONG has higher throughput than IVFPQ on GPU included in FAISS \cite{johnson_billion-scale_2021} and HNSW on CPU.}
Similarly, GANNS is a GPU-friendly graph search and construction method tailored for NSW, HNSW, and $k$ nearest neighbor graphs on GPUs \cite{yu_gpu-accelerated_2022}. This approach further advances the efficiency of graph-based ANNS on GPU architectures by modifying data structures for GPUs and reducing their operation time.
Additionally, Groh et al. present GGNN, a fast graph construction and search method designed explicitly for GPUs \cite{groh_ggnn_2022}.
Their work also improves the overall performance of ANNS on GPUs by improving data structures for GPUs and utilizing fast shared memory.
Some studies have pointed out and managed the issue that graph construction can be time-consuming by using the advantage that a proximity graph can be reused once it is constructed. Unfortunately, there still remains a critical challenge in efficiently designing proximity graphs that are well-suited for GPU architectures in both construction and search.
Despite the progress made in using GPUs for graph-based ANNS, most existing studies focus on adapting or optimizing existing graphs for GPU utilization rather than specifically designing proximity graphs from the ground up to fully leverage the GPU's capabilities in both graph construction and subsequent search operations. 

This paper proposes 1) a proximity graph suitable for parallel computing in graph construction and query search and 2) a fast ANNS graph construction and search implementation, {\bf CAGRA} (Cuda Anns GRAph-based), optimized for NVIDIA GPU. Our graph is search implementation-centric and designed to increase the efficiency of a massively parallel computing device, like the GPU, rather than theoretical graph quality.

\changed{The summary of our contributions is as follows:
\begin{itemize}
    \item We propose a proximity graph for ANNS and its construction method suitable for massively parallel computing.
    This method reduces the memory footprint and usage, which can be the performance bottleneck in the graph optimization process, by avoiding exact similarity computation.
    \item We provide a search implementation optimized for GPU, which is designed to gain high throughput in both single and large-batch queries.
    We harness software warp splitting and forgettable hash table management to utilize the GPU resource efficiently.
    \item We demonstrate that CAGRA achieves a higher performance in graph construction and query than the state-of-the-art graph-based ANNS implementations for CPU and GPU.
    In graph construction time, CAGRA is $2.2 \textendash{} 27\times$ faster than HNSW.
    In large-batch query throughput in the 90\% to 95\% recall range, CAGRA is $33\textendash{} 77\times$ faster than HNSW and is $3.8\textendash{} 8.8\times$ faster than the SOTA implementations for GPU.
    For a single query, CAGRA is $3.4\textendash{} 53\times$ faster than HNSW at 95\% recall.
\end{itemize}
}

\section{Background}
\subsection{Approximate Nearest Neighbor search}
In the $k$-NNS problem, we obtain $k$ vectors $x_{i_1}, x_{i_2}, \ldots{}, x_{i_k} \in \mathbb{R}^{n}$ that satisfy the following condition from a dataset $\mathcal{D} \in \mathbb{R}^{n \times N}$:
\begin{equation}
    i_1, i_2, \ldots{}, i_k = k\text{-}\mathrm{argmin}_i \operatorname{Distance}\left(x_i, q\right),
\end{equation}
where $q \in \mathbb{R}^{n}$ is a given query vector, $k\text{-}\mathrm{argmin}_i$ is the top-$k$ arguments in ascending order, and $\operatorname{Distance}(\cdot)$ ($\mathbb{R}^n \times \mathbb{R}^n \rightarrow \mathbb{R}$) is a distance measure.
The distance is typically the L2 norm or cosine similarity.
Although $k$-ANNS obtains $k$ similar vectors to a given query vector, the results are not always exact.
We evaluate the accuracy of the results for a query as recall:
\begin{equation}
    \mathrm{recall} = |U_\text{ANNS} \cap U_\text{NNS}| / |U_\text{NNS}|,
\end{equation}
where $U_\text{ANNS}$ is the set of resulting vectors obtained by ANNS and $U_\text{NNS}$ is by NNS.
We denote the recall of $k$-ANNS as ``recall@$k$''.
There is typically a trade-off between the recall and throughput (QPS; Query Per Second).

\subsection{CUDA}

The GPU, or Graphics Processing Unit, has been broadly used for general-purpose computing in recent years, whereas it was initially developed strictly for graphics processing. 
NVIDIA proposes and has been developing CUDA, which allows us to leverage the high-performance computing capability of the GPU for general-purpose computing.
While the GPU has higher parallelism and memory bandwidth than the CPU, we have to be able to abstract parallelism from an algorithm and map it to the architecture of the GPU to leverage its high performance.
Therefore, not all applications can gain higher performance by just using the GPU.
\revised{
We briefly introduce the architecture of NVIDIA GPU in the following sections, and further information on the architecture can be found in \cite{nvidia_cuda_2023}.
}

\subsubsection{Thread hierarchy}
\label{sec:cuda-thread}
In the NVIDIA GPU thread hierarchy, 32 threads in a group called ``warp'' execute the same instruction simultaneously.
On the other hand, different instructions are not executed in a warp in parallel, leading to a performance degradation called ``warp divergence''.
A group of up to 32 warps composes a CTA (Cooperative Thread Array), or thread block, and a CTA is executed on a single GPU streaming multiprocessor (SM).
The SM is like a core in a multi-core CPU, and since there are many SMs on recent GPUs, we can launch and operate multiple CTAs at a time.

\subsubsection{Memory hierarchy}
In the memory hierarchy of the GPU, the device memory has the largest size, and all threads can access the same memory space.
Shared memory, on the other hand, is a local memory used within each CTA and all the threads in the CTA share the memory space.
While the size of shared memory is much smaller than the device memory, it has lower latency and higher bandwidth.
Registers are local data storage for each thread and have lower latency than the shared memory.


\subsection{Related work}
Various algorithms for graph-based ANNS were introduced in Sec. \ref{sec:introduction} so we focus this section specifically on highly parallel and GPU-accelerated graph-based ANNS implementations whilst outlining how our contributions compare to them.
\subsubsection{SONG}
SONG is the first graph-based ANN implementation for GPU proposed by Zhao \textit{et al.} \cite{zhao_song_2020}.
Unlike CAGRA, this method does not contribute a faster graph construction technique and relies on other methods like NSW \cite{malkov_approximate_2014}, NSG \cite{fu_fast_2019}, and DPG \cite{li_approximate_2020}.
SONG proposes a dataset and several optimizations for the GPU, in which they use open address hash table, bounded priority queue, and dynamic allocation reduction.
They have achieved 10--$180 \times$ speed up on the GPU compared to single-threaded HNSW on CPU.
\subsubsection{GGNN}
GGNN is a GPU-friendly implementation of graph-based ANNS proposed by Groh \textit{et al.} \cite{groh_ggnn_2022} that, like CAGRA, provides both a high-throughput search implementation and a fast graph construction technique that can utilize high parallelism.
GGNN was demonstrated to outperform SONG in large-batch searches.
\subsubsection{GANNS}
GANNS is also a GPU-accelerated graph construction and search method proposed by Yu \textit{et al.} \cite{yu_gpu-accelerated_2022}.
They propose a GPU-based NSW graph construction method and show that both the proximity graph construction and search performance are better than SONG.

Unfortunately, between GGNN and GANNS, it is unclear which is the state-of-the-art graph-based ANNS GPU implementation since there is no study on comparison between them. In addition, all of the above GPU implementations are focused on applications with a large number of queries. To the best of our knowledge, no GPU implementation outperforms the CPU implementation for applications with small-batch queries. In this paper, we will show that CAGRA outperforms both the CPU and GPU in graph construction and search. 

\section{CAGRA graph}
\begin{figure*}[t]
    \centering
    \includegraphics[width=\textwidth]{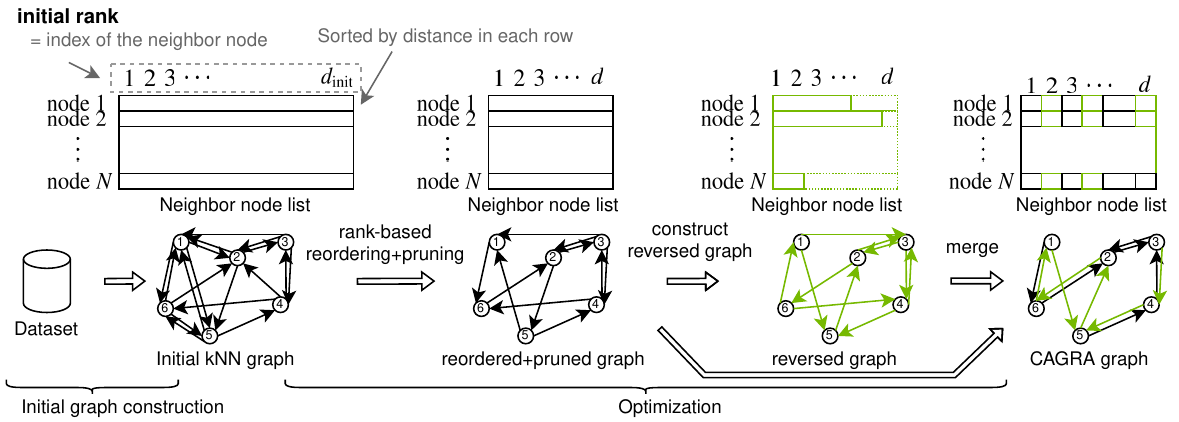}
    \caption{Construction flow of the CAGRA graph.}
    \label{fig:cagra-graph}
\end{figure*}
%
In this section, we explain the design and features of the CAGRA graph. While some graphs are designed to follow or approximate graphs with some theoretical properties, including monotonicity, the CAGRA graph is a search implementation-centric and heuristically optimized graph.
The CAGRA graph has the following features:
\begin{itemize}
    \item \textbf{Fixed out-degree} ($d$).
        By fixing the out-degree, we can utilize the massive parallelism of GPU effectively.
        Most graph-based ANNS algorithms build and utilize non-fixed out-degree graphs.
        In single-thread execution on a CPU, a non-fixed degree for each node has an advantage in that we can reduce the less important distance computation by keeping only essential edge connections.
        However, in the case of GPU, too small a degree doesn't fully saturate the computing resources allocated to each CTA, leading to lower hardware utilization.
        Rather, it is better to expand the search space using all the available compute resources, as it won't increase the overall compute time.
        Another advantage is that fixing the degree allows more uniform computation, thus creating less load imbalance during the parallel graph traversal phase, which would lead to low hardware utilization.
        \changed{We set the degree depending on the dataset and required recall and throughput.}
    \item \textbf{Directional}.
        Since the out-degree is fixed, the graph is naturally directional.
    \item \textbf{No hierarchy}.
        HNSW, for instance, employs a hierarchical graph to determine the initial nodes on the bottom layer.
        However, in the case of GPU, we can obtain compatible initial nodes by randomly picking some nodes and comparing their distances to the query, thus employing the high parallelism and memory bandwidth of GPU.
        The detail of the search algorithm is in Sec. \ref{sec:search}.
\end{itemize}

Two main steps are involved in the construction of the CAGRA graph: 1) building a $k$-NN graph and 2) optimizing it to enhance search recall and throughput.
We chose $k$-NN graph as the base graph because the fixed out-degree graph is well-suited for efficient GPU operations, and we can rapidly build it using nearest neighbors descent (NN-descent) \cite{wang_comprehensive_2021} even on the GPU \cite{wang_fast_2021}.
The following section outlines the heuristic graph optimization approach and its parallel computation algorithm. 

\subsection{Graph optimization}
The primary objective of CAGRA graph optimization is to enhance reachability among a large number of nodes.
Reachability is characterized by two properties: 1) whether we can traverse from any arbitrary node to another arbitrary node and 2) the number of nodes we can traverse from one node within a specified number of path traversals.

To assess property 1), we measure the number of strongly connected components (CC) in the graph.
The number of CC is determined as follows:

There is no guarantee that the base kNN graph is not disconnected \cite{nolet_cuslink_2023}, and the weak CC represents the number of subgraphs in the graph.
Additionally, in the case of a directional edge graph, there may be scenarios where traversal from one node to another is not possible, even if the graph is not entirely disconnected.
A graph is considered strongly connected when an arbitrary node in the graph can reach any other node.
The number of strong CC is the count of node groups in the graph, where each group forms a strongly connected subgraph.
A smaller number of strong CC are preferred because a larger number of CC can lead to more unreachable nodes starting from a search start node.

To assess property 2), we utilize the average 2-hop node count ($N_{2\text{hop}}$) for all nodes in the graph as the metric.
The 2-hop node count of a given node is defined as the number of nodes that can be reached in two steps from the node.
Its maximum value is determined as $N_{2\text{hop}}^\text{max} = d + d^2$ where $d$ is the degree of the graph.
A higher average 2-hop node count indicates that more nodes can be explored during specific search iteration steps.

In the CAGRA graph optimization, two key techniques are employed on the initial kNN graph: {\bf reordering} and {\bf reverse edge addition}.
Reordering is a technique that reorders each edge of the initial kNN graph in an order that increases the diversity of the graph, rather than in the order of its length, and has the primary effect of increasing 2-hop node counts.
Reverse edge addition is a technique often used in other graph-based ANN implementations and improves node reachability while reducing strong CC values.

\subsection{Graph construction and optimization algorithm}
The CAGRA graph is a directed graph where the degree, $d$, of all nodes is the same.
The construction of the graph consists of two stages: 1) construction of an initial graph and 2) optimization, as shown in Fig. \ref{fig:cagra-graph}.

\subsubsection{Initial graph construction} 
We construct a $k$-NN graph as an initial graph where the degree of each node is $k=d_\text{init}$.
We use NN-Descent \cite{dong_efficient_2011} to construct the graph and will typically set $d_\text{init}$ to be $2d$ or $3d$, where $d$ is the degree of the final CAGRA graph.
As a final step in this process, we sort the connected node list of each node in ascending order based on distance from the source node.
This sorting operation can be efficiently executed in parallel using GPUs since no dependency exists in the computation for each individual node list.
\changed{
We assume that the initial $k$-NN graph has sufficient connectivity among nodes.
And the goal of the CAGRA graph optimization is to reduce the degree of the graph to reduce the size while preserving the reachability.
}

\subsubsection{Graph optimization}

\begin{figure}[t]
    \centering
    \includegraphics[width=\linewidth]{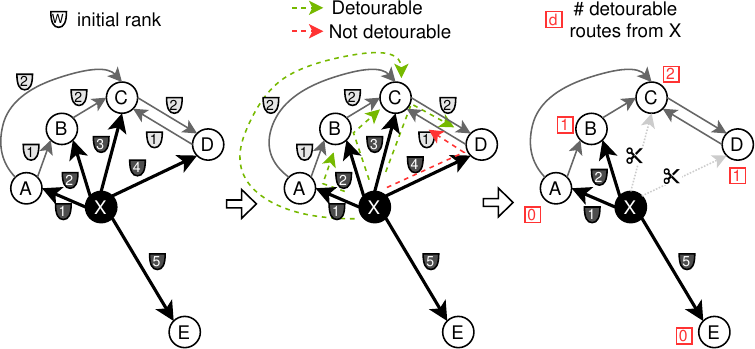}
    \caption{CAGRA edge reordering and pruning flow. We assume pruning edges from the node $X$.
    {\bf Left}: The initial rank of the edges from $X$ and other related edges.
    {\bf Middle}: Possible two-hop routes, classified as detourable and not detourable by Eq. \ref{eq:detourable-condition}. We use the rank instead of the distance.
    {\bf Right}: The number of detourable routes of each node connected to $X$. The edges are discarded from the end of the list ordered by the number of detourable routes.
    In this example, the nodes $A$, $B$, and $E$ are preserved as the neighbors of node $X$, although the node $E$ is the farthest in the initial neighbors of node $X$ in the distance.
    }
    \label{fig:caga-prune}
\end{figure}
The optimization process involves two steps: 1) reordering edges, and 2) adding reverse edges.
It takes the initial graph as input and produces the CAGRA graph as output.
This process offers notable advantages: i) it no longer requires the dataset or distance computation, and ii) it allows for many computations to be executed in parallel without complex dependencies.

\changed{
In the reordering edges step, we determine the significance of an edge, {\bf rank}, to prune the edges at the end of the entire optimization.
Existing pruning algorithms prune an edge from one node to another if it can be bypassed using another route (detourable route) that satisfies certain criteria.
For instance, in NGT \cite{iwasaki_optimization_2018}, it defines the detourable route from $X$ to $Y$ as a pair of two edges as follows:
\begin{equation}
    \label{eq:detourable-condition}
    (e_{X\rightarrow Z}, e_{Z\rightarrow Y}) \text{ s.t. } \operatorname{max}(w_{X\rightarrow Z}, w_{Z \rightarrow Y}) < w_{X \rightarrow Y},
\end{equation}
where $e_{\cdot\rightarrow\cdot}$ and $w_{\cdot\rightarrow\cdot}$ denote a directed edge between two nodes and the distance, respectively, and $Z$ is a node with a direct connection from $X$ and a direct connection to $Y$.
Based on this pruning, we consider two reordering techniques, {\bf distance-based} and {\bf rank-based} reordering, and adopt rank-based in the CAGRA graph optimization.
}
\revised{
While the complexity of both the reordering operations is $\mathcal{O}(Nd^3)$, the distance-based strategy requires distance calculations between one node and its neighbors.
}

\changed{
In distance-based reordering, we first count the number of detourable routes for each edge and reorder the node list by the counts in ascending.
A smaller number of detourable routes for an edge indicates that this edge is more important to keep the 2-hop node counts.
Then, we set the position of an edge in the node list as a rank of the edge, which is an indicator of the importance of the edge.
The computation of distance-based reordering has high parallelism since we can count the detourable routes for each edge in parallel.
However, we need to compute the distances on the fly during the operation or make a distance table before the operation, making this method impractical for a large dataset.
In the former case, we need $N \times d_\text{init} \times (d_\text{init}-1)$ distance computations, and in the latter case, we need a distance table with $N \times d_\text{init}$ entries on memory, where $N$ is the size of the dataset.
}

\changed{
For rank-based reordering, we set the position of the edge in the neighbor node list, which is sorted by distance at the end of the initial graph construction, as the initial rank, similar to distance-based reordering.
Then, we reorder the edges in the same way as distance-based reordering, but we use the initial rank instead of the distance, as shown in Fig. \ref{fig:caga-prune}.
In other words, we approximate the distance by the initial rank.
This approximation allows us not to compute the impractical amount of distance computations and not to store the large size of the distance table in memory.
We set the order index of a node as the rank, the same as distance-based.
We adopt rank-based reordering in the CAGRA graph optimization and only keep top $d$ neighbors for each node (pruning).
}

After reordering, we create a reversed graph where all edges have opposite directions of the reordered and pruned graph.
Since the number of incoming edges per node, or in-degree, is not fixed in the reordered graph, the out-degree of each node in the reversed graph is also not fixed.
However, we set the upper bound of the degree to $d$ because of an attribute of the next operation.
And we make the reversed graph so that the reversed edges are sorted by the rank in the pruned graph in ascending order.
It means ``{\it Someone who considers you are more important is also more important to you}''.
\revised{The process of adding reverse edges has a complexity of $\mathcal{O}(Nd)$.}

Finally, we merge the pruned graph and reversed graph.
In this process, we basically take $d/2$ children for each parent node from each graph and interleave them.
When the number of children for a parent node in the reversed edge graph is fewer than $d/2$, we compensate them by taking from the pruned graph.

\revised{
To find an optimal $d$, we build graphs with different numbers, such as 32, 64, and 96, and measure their search performance.
There is no deterministic way to find the parameter in a single shot since it depends on the dataset and user requirements. This is not unlike the hyper-parameters ofother methods, for example, the maximum out-degree of the HNSW (libhnsw) graph.
Increasing the out-degree improves the recall while the search throughput degrades.
}

\subsection{Evaluation of the CAGRA graph and optimization}
This section reveals the following question:
\begin{enumerate}[label={\bf Q-A\arabic*},leftmargin=3em]
    \item How much do the CC and 2-hop node counts improve with the CAGRA graph optimization?
    \item How fast is rank-based reordering compared to distance-based?
    \item Does rank-based reordering have compatible recall with distance-based?
\end{enumerate}

We use the datasets in Table \ref{tab:datasets}.
\begin{table}[t]
\caption{Datasets used in the evaluations}
\label{tab:datasets}
\begin{tabular}{l|lll|l}
                                         & Dim ($n$) & Size ($N$) & Data type & \begin{tabular}[c]{@{}l@{}}CAGRA graph\\ degree ($d$)\end{tabular} \\ \hline
SIFT-1M $^1$                  & 128       & 1M         & float     & 32                                                                 \\
GIST-1M $^1$  & 960       & 1M         & float     & 48                                                                 \\
GloVe-200 \cite{pennington_glove_2014}   & 200       & 1183514    & float     & 80                                                                 \\
NYTimes\cite{aumuller_ann-benchmarks_2020}                                  & 256       & 290K       & float     & 64                                                                 \\
DEEP-1M $^2$     & 96        & 1M         & float     & 32                                                                 \\
DEEP-10M $^2$     & 96        & 10M        & float     & 32                                                                 \\
DEEP-100M \cite{yandex_efficient_2016}   & 96        & 100M       & float     & 32 
\end{tabular}
\end{table}
\footnotetext[1]{Datasets for ANNS: \url{http://corpus-texmex.irisa.fr/}}
\footnotetext[2]{First 1M and 10M vectors of DEEP-100M dataset \cite{yandex_efficient_2016}.}
All experiments are conducted on a \revised{DGX A100 server equipped} with AMD EPYC 7742 CPU (64 cores) and NVIDIA A100 (80GB) GPU, \revised{high-end processors released within a similar timeframe}.
We put both the dataset and graph on the device memory of the GPU.
\revised{In the CAGRA graph construction, we build an initial $k$-NN graph on GPU and optimize it on CPU.}

\subsubsection{{\bf Q-A1:} Connected components and 2-hop node count}
In the CAGRA graph construction, two optimizations are performed on the initial $k$-NN graph: reordering and reverse edge addition.
Then, \textit{how much effect does each optimization have?}
To evaluate this, we have compared the properties of a standard $k$-NN graph, a partially optimized CAGRA graph (using only one optimization), and a fully optimized CAGRA graph (using both optimizations).


The results of the 2-hop node count and the strong CC experiments are shown in Fig. \ref{fig:cagra-opt-effect}.
In the case of the 2-hop node count, we observe that both optimizations increase the average 2-hop count, and the effect of the reordering is more significant compared to the reverse edge addition.
The results also show that reverse edge addition significantly affects the strong CC more than reordering.

\begin{figure}[t]
    \centering
    \includegraphics[width=\linewidth]{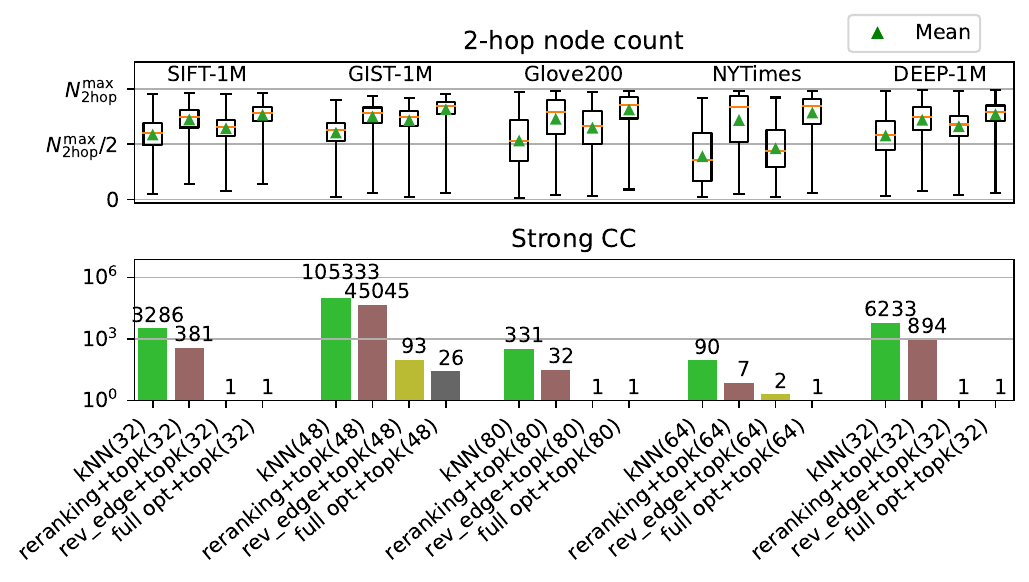}
    \caption{
    The 2-hop node counts and strong CC comparison among a $k$-NN graph, partially and fully optimized graphs by CAGRA from an initial $k$-NN graph.
    The number in each bracket in the label is the degree of the graph ($d$), and we set the degree of the initial graph as $d_\text{init}=3d$.
    }
    \label{fig:cagra-opt-effect}
\end{figure}

\subsubsection{{\bf Q-A2:} The reordering method's advantage on compute time}
In the reordering optimization, we avoid the distance computation altogether, reducing the computational overhead, and leading to faster optimization.
So then, \textit{how does that improve the total optimization time?}
We have measured the optimization time, as shown in Fig. \ref{fig:norm-based-construction-time}.
The rank-based CAGRA optimization is faster than the distance-based for all datasets by as much as $1.9\times$.
Furthermore, while we were still able to perform the rank-based optimization, we experienced an out-of-memory error that prevented us from performing the distance-based optimization on DEEP-100M.
These results show that rank-based optimization is faster than distance-based and supports larger datasets.

\begin{figure}[t]
    \centering
    \includegraphics[width=\linewidth]{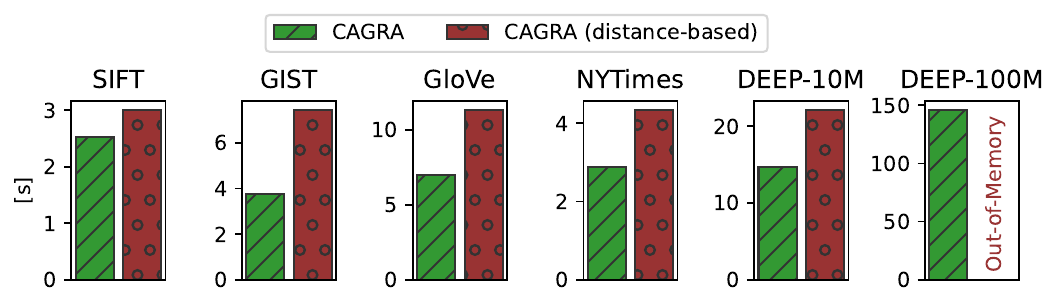}
    \caption{CAGRA graph optimization time comparison with rank- and distance-based reordering.}
    \label{fig:norm-based-construction-time}
\end{figure}

\subsubsection{{\bf Q-A3:} Seach performance comparison to distance-based optimization}
\begin{figure}[t]
    \centering
    \includegraphics[width=\linewidth]{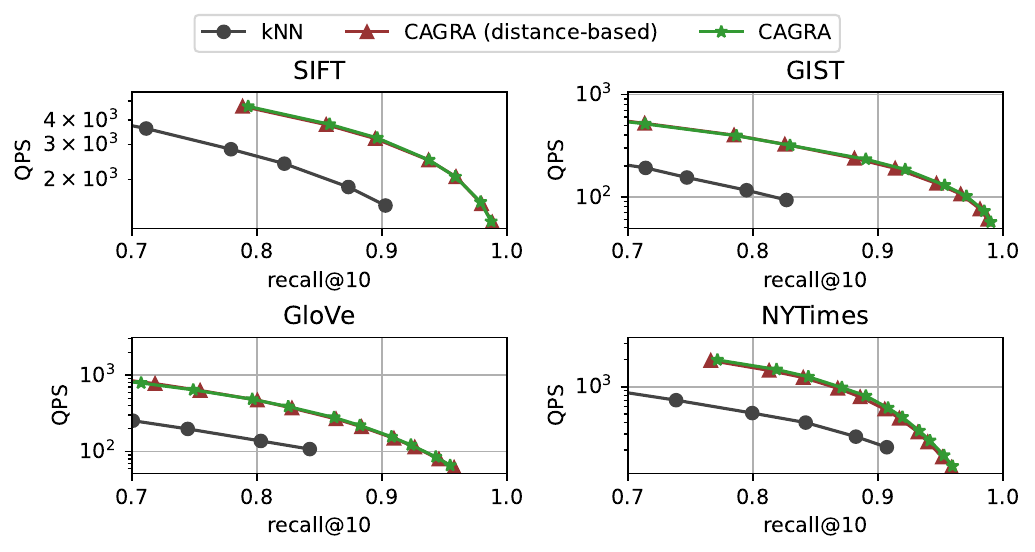}
    \caption{
    CAGRA search performance comparison between the graphs optimized by rank- and distance-based reordering.
    CAGRA performs rank-based reordering while CAGRA (distance-based) performs distance-based. 
    }
    \label{fig:norm-based-search-performance}
\end{figure}
The recall that a graph can potentially achieve and the number of iterations to obtain a specific recall will vary by the graph construction methods, including the reordering priority criteria in the CAGRA graph optimization.
In CAGRA, we reduce the graph optimization time, avoiding distance computation and instead using rank as the priority criteria.
So then, \textit{does the CAGRA graph have the compatible search performance compared to distance-based optimization?}
To answer this question, we have tested both rank-based and distance-based reordering during CAGRA graph optimization and measured the throughput and recall of a query search process using the graph, as shown in Fig. \ref{fig:norm-based-search-performance}.
This confirms the recall-throughput balance is almost the same while the rank-based graph construction time is shorter, as demonstrated in Q-A2.

\section{CAGRA Search}
\label{sec:search}
In this section, we explain CAGRA's search algorithm and how we map it onto the GPU.
\subsection{Algorithm}
\begin{figure}[t]
    \centering
    \includegraphics[width=0.8\linewidth]{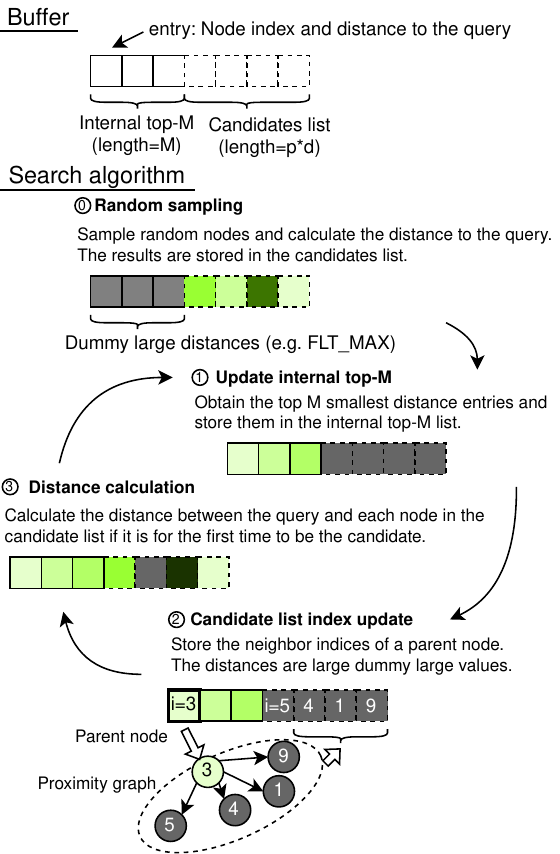}
    \caption{\textit{Top}: The buffer layout used in the CAGRA search. \textit{Bottom}: The algorithm of the CAGRA search.}
    \label{fig:cagra-search-algorithm}
\end{figure}
The CAGRA search algorithm uses a sequential memory buffer consisting of an {\bf internal top-$M$ list} (typically known as a priority queue in other algorithms) and its {\bf candidate list}, as shown at the top of Fig. \ref{fig:cagra-search-algorithm}.
The length of the internal top-$M$ list is $M (\geq k)$, and the candidate list is $p \times d$, where $p$ is the number of source nodes of the graph traversed in each iteration, and $d$ is the degree of the CAGRA graph.
Each buffer element is a key/value pair containing a node index and the corresponding distance between the node and the query.

As shown at the bottom of Fig. \ref{fig:cagra-search-algorithm}, The search calculation is as follows:
\begin{enumerate}[label=\protect\circled{\arabic*}]
\setcounter{enumi}{-1}
    \item {\bf Random sampling} (initialization step): We choose $p \times d$ uniformly random node indices using a pseudo-random number generator and compute the distance between each node and the query.
    The results are stored in the candidate list.
    We set the internal top-$M$ list with dummy entries where the distance values are large enough to be the last in the next \circled{1} sorting process.
    For instance, if the distance is stored in {\tt float} data type, {\tt FLT\_MAX}.
    \item {\bf Internal top-$M$ list update}: We pick up top-$M$ nodes with the smallest distance in the entire buffer and store the results in the internal top-$M$ list.
    \item {\bf Candidate list index update} (graph traversal step): We pick up all neighbor indices of the top-$p$ nodes in the internal top-$M$ list where they have not been parents.
    The result node indices are stored in the candidate list.
    This step does not calculate the distance between each node in the candidate list and the query.
    \item {\bf Distance calculation}: We calculate the distance between each node in the candidate list and the query only if this is the node's first time being in the candidate list for the query.
    This conditional branch prunes unnecessary computations since distances don't need to be recomputed if they were already computed in a previous iteration. For instance, if a node has already been in the list and the distance is
    \begin{itemize}
        \item small enough to stay in the top-$M$ list, then it should already be in the list.
        \item large enough not to be in the top-$M$ list, then it should not be added again.
    \end{itemize}
\end{enumerate}
We process \circled{1} $\sim$ \circled{3} iteratively until the index numbers in the top-$M$ list converge, meaning they remain unchanged from the previous iteration.
Finally, we output the top-$k$ entries of the internal top-$M$ list as the result of ANNS.

\subsection{Elemental technologies and designs}
This section explains the elemental technologies and designs we use in the CAGRA search implementation on GPU.

\subsubsection{Warp splitting}
\label{sec:team-size}
As described in Sec. \ref{sec:cuda-thread}, a warp consists of 32 threads that execute the same instruction simultaneously, representing the smallest parallel thread group in the hardware.
In the CAGRA search implementation, we introduce a software-level division of the warp into even smaller thread groups, referred to as {\bf teams}. Each team consists of a specific number of threads, which we term the {\bf team size}.

This division allows us to enhance GPU utilization for the following reasons: Consider the latency of device memory load, where a $128$-bit load instruction is the most efficient.
We typically map one distance computation to one warp and utilize warp shuffle instructions to compute it across the threads collaboratively.
However, when the dataset dimension is $96$, and the data type is {\tt float} ($4$-byte), the total bit length of a dataset vector is $3072$ bits, which is smaller than the bit length loaded when all $32$ threads in a warp issue the $128$-bit load instruction, $4096$ bits.
Consequently, this will leave some threads in the warp which do not issue the load instruction, resulting in inefficient GPU usage.
Now consider mapping one distance computation to a team with a team size of $8$, and one team can load $1024$ bits in one instruction.
This allows the entire vector to be loaded by repeating the loading three times in all threads of the team.
Additionally, the other teams within the same warp can calculate the distances between the query and the other nodes in the candidate list, thereby maximizing GPU utilization and efficiency. Although we split the warp into teams in software, we don't encounter warp divergence since all of the teams in each warp still execute the same instructions.

\subsubsection{Top-$M$ calculation}
In \circled{1}, we obtain top-$M$ distance entries from the buffer.
Since we can assume that the individual internal top-$M$ list parts have already been sorted, we can reduce the computation compared to the full top-$M$ computation to the buffer.
More specifically, we first sort the candidate buffer and merge it with the internal top-$M$ buffer through the merge process of the bitonic sort \cite{johnson_billion-scale_2021}.
We use the single warp-level bitonic sort when the candidate buffer size is less or equal to $512$, while we use a radix-based sort using within a single CTA when it is larger than $512$.
This design is based on the observation that when the candidate buffer is small enough, we can quickly sort the candidate list right in the registers of a single warp without the shared memory footprint, while we need to use the shared memory when the list length is large, resulting in a performance degradation compared to the radix-based sort.

\subsubsection{Hash table for visited node list management}
\label{sec:hash}
In \circled{3}, we calculate the distance between the query and each node in the candidate list only the first time the node appears in the list. This requires a mechanism for recording whether a node has been in the list before and, in a similar manner to the SONG algorithm  \cite{zhao_song_2020}, we use an open addressing hash table to manage the visited node list in the CAGRA search.

The number of potential entries in the hash table is calculated as $I_\text{max} \times p \times d$, where $I_\text{max}$ represents the maximum number of search iterations. We set the hash table size to at least twice this value to reduce the likelihood of hash collisions.

If we place the hash table in limited memory, such as shared memory, and it exceeds the memory's capacity, we use a smaller hash table with periodic \textit{table resetting}, meaning the hash table evicts previously visited nodes at certain intervals. After resetting the table, we only register the nodes present in the internal top-$M$ list to the hash table at that moment.
Although this process may increase the number of distance computations, catastrophic recall degradation will not occur, as mentioned in \cite{zhao_song_2020}. We refer to this type of hash table management, which is meant for limited memory and uses table resetting, as {\bf forgettable hash table management}.
We set the number of entries of the hash table as $2^8 \sim 2^{13}$ and the reset interval as typically $1 \sim 4$ depending on the graph and search parameters $M, d,$ and $p$.
This hash table management reduces the shared memory usage per query, \revised{typically $\leq$ 4 kB,} resulting in higher parallel efficiency in a large-batch query \revised{on almost all generation of NVIDIA GPUs.}

\subsubsection{1-bit parented node management}
In step \circled{2}, we select a specific number of nodes that have not previously been parents and assign their neighbor indices to the candidate list.
To keep track of whether a node has acted as a parent, we utilize the Most Significant Bit (MSB) of the index variable in the buffer as a flag for recording this information.
An alternative approach could involve using another hash table, but we choose not to adopt it due to a latency disadvantage.
We need to search the entry in the hash table if we use the alternative method, while we can check whether a node has acted as a parent just by reading the MSB of the node index in the list.
This method comes with a disadvantage, however, as it imposes a limitation on the size of the dataset and restricts it to half of the maximum value representable by the index data type.
For instance, when the data type used is {\tt uint32\_t}, the supported maximum size of the dataset is only $2^{31}-1$, compared to the maximum value of $2^{32}-1$ if we didn't utilize the MSB for this flag.

\subsection{Implementation}
\label{sec:search-imp}

\begin{table}[]
\caption{The summary of the difference between the single- and multi-CTA modes.}
\label{tab:search-mode}
\begin{tabular}{l|ll}
                      & single-CTA    & multi-CTA                                                                          \\ \hline
\rowcolor[HTML]{e0e0e0} 
Use case              & large-batch   & \begin{tabular}[c]{@{}l@{}}small-batch or\\ higher recall is required\end{tabular} \\
per 1 query           & single CTA    & multiple CTA                                                                       \\
\rowcolor[HTML]{E0E0e0} 
Hash table location   & Shared memory & Device memory                                                                      \\
Hash table management & Forgettable   & Standard                                                                          
\end{tabular}
\end{table}

This section explains the features and optimizations of the CAGRA search implementation on the GPU.
The CAGRA search contains separate implementations for \textbf{single-CTA} and \textbf{multi-CTA}. While the basic search strategy and operations are the same in both implementations, mapping the hardware to the queries differs. The summary of these implementations is shown in Table \ref{tab:search-mode}.

\subsubsection{Single-CTA implementation}
\label{sec:single-cta}
As the name implies, the single-CTA implementation is designed to efficiently process queries by mapping each query to one CTA. The target batch size for this implementation ranges from middle to large values, such as $100$ and above.
Leveraging the parallel processing capabilities of GPUs, multiple of these single-CTAs can be executed simultaneously, enabling efficient handling of multiple queries and effective GPU resource utilization. However, we note that relatively small batch sizes can leave the GPU resources underutilized, leading to suboptimal performance.

To implement the single-CTA approach, we have developed a kernel function that handles the entire search process \circled{0} $\sim$ \circled{3}, placing the hash table in shared memory rather than device memory. As part of our optimization exploration, we have also considered an alternative implementation involving separate kernel functions, with each function handling a specific step of the search process (\circled{0} $\sim$ \circled{3}).
However, extensive testing revealed that the overhead of launching multiple kernels outweighs any potential performance gains. As a result, we have concluded that adopting the multi-kernel approach is not advantageous, and we instead prefer our single-kernel implementation.

In the throughput analysis of the implementation, we have observed that the memory bandwidth of the device memory limits the performance of the kernel function when the query batch size and the dimension of the dataset are large.
We propose an approach involving low-precision data types for dataset vectors to address the memory bandwidth limitations and enhance overall throughput.
By reducing the memory footprint, this optimization technique aims to expedite data transfer between memory and processing units, thereby increasing throughput.
However, it is crucial to carefully assess the trade-off between throughput enhancement and the potential impact on recall.

\subsubsection{Multi-CTA implementation}
\label{sec:multi_cta_impl}
The multi-CTA implementation is designed to map one query to multiple CTAs, with a target batch size typically ranging from small values, such as $1 \sim 100$.
In contrast to the single-CTA implementation, which launches only as many CTAs as the query batch size, the multi-CTA approach maximizes GPU resource utilization by employing multiple CTAs to process a single query.
As a result, this implementation achieves higher GPU utilization and enhances query processing efficiency, even when dealing with small batch sizes.


In this implementation, the hash table is stored in device memory as it needs to be shared with multiple CTAs.
Each CTA traverses graph nodes, managing its own internal top-$M$ list and candidate list by setting the number of parent nodes as $1$ while sharing the hash table.
Therefore, while we search up to $p \times d$ nodes in each iteration in single-CTA, we search up to \textit{the number of CTA we launch} $\times d$ nodes in each iteration.
Since we typically set $p=1$ to maximize the throughput of single-CTA, the number of nodes visited in each iteration in multi-CTA is larger than in single-CTA, leading to higher recall if the number of iterations is the same.
 
We explored an alternative approach involving graph-sharding, which is commonly used for multi-node ANNS computations \cite{johnson_billion-scale_2021}, with the goal of maximizing GPU resource utilization for small batch sizes.
However, we decided against it for practical reasons that pose challenges to execution optimizations, such as the reliance on specific graph structures to create the subgraphs, as well as the shuffling and splitting of the indices to create sub-datasets of the target dataset.

Subsequently, we independently built graphs for each sub-dataset, similar to the graph construction method used in GGNN \cite{groh_ggnn_2022}.
During the search phase, assigning each sub-graph to a single CTA did result in high GPU resource utilization when the number of sub-datasets was sufficiently large, however, despite its potential advantages, this method presented several issues.
For example, determining the optimal number of splits depends on factors such as the query batch size and the hardware configuration, specifically the number of SMs on the GPU.
Creating a series of sub-graphs for each batch size and GPU configuration is not feasible in practice, and it makes this approach impractical.
Nevertheless, we recognize that the sharding technique could be well-suited for extending graph-based ANNS to a multi-GPU environment, where each GPU is assigned to process one sub-graph independently.

\subsubsection{Implementation choice}
\begin{figure}[t]
    \centering
    \includegraphics[width=0.8\linewidth]{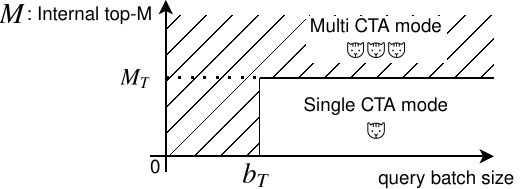}
    \caption{The rule to choose a suitable CAGRA search implementation.}
    \label{fig:implementation-choose}
\end{figure}
As mentioned above, we have two implementations targeting small and large batch sizes and we select the implementation based on both the batch size and the internal top-$M$ size, as shown in Fig. \ref{fig:implementation-choose}. We use the multi-CTA implementation when the query batch size is smaller than a threshold $b_T$ or when the internal top-$M$ size is larger than a threshold $M_T$, since the computing cost of \circled{1} is large in the single-CTA implementation for these cases, increasing the computing time.
When the multi-CTA implementation is not used, we fall back to the single-CTA implementation.
While the proper thresholds depend on the hardware, we recommend $M_T=512$ and $b_T=$ ``the number of SMs on the GPU'' empirically.

\subsection{Evaluation of the CAGRA search implementation}
This section reveals the following question:
\begin{enumerate}[label={\bf Q-B\arabic*},leftmargin=3em]
    \item How much effect does the warp splitting have on the throughput?
    \item How much effect does the forgettable hash have on the throughput?
    \item Which cases single-CTA is faster than multi-CTA?
\end{enumerate}

\subsubsection{{\bf Q-B1:} The effect of team size in throughput}
\begin{figure}[t]
    \centering
    \includegraphics[width=\linewidth]{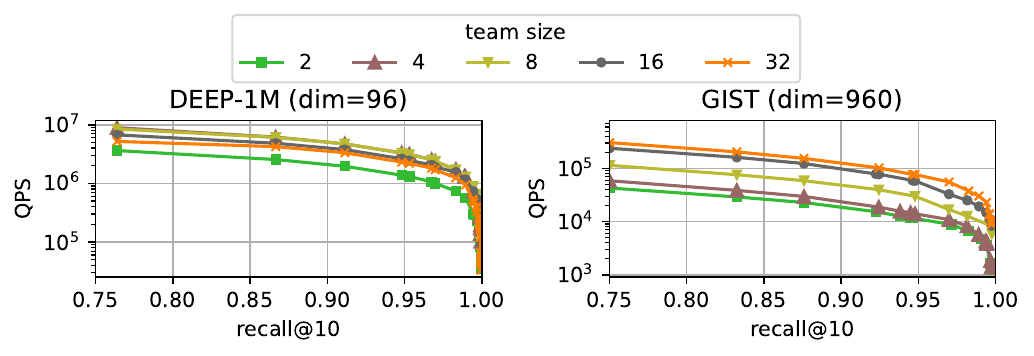}
    \caption{Search performance comparison among different team sizes.}
    \label{fig:team-size}
\end{figure}
We split the warp into multiple teams in software to efficiently utilize the GPU resources, as mentioned in Sec. \ref{sec:team-size}.
Then, \textit{how much effect does this warp splitting have?}
\changed{We compare the performance among different team sizes for the DEEP-1M and GIST datasets in Fig. \ref{fig:team-size}.
In the evaluation result for DEEP-1M, a relatively small dimension dataset, we can gain the highest performance when the team size is $4$ or $8$ while maintaining recall.
When the team size is too small, such as $2$, the number of registers per thread becomes too large, leading to performance degradation.
On the other hand, in the search performance for GIST, a relatively large dimension dataset, we achieved the highest performance when the team size was $32$.
In this case, we can utilize the GPU resources efficiently even if we do not split the warp.
Instead, decreasing the team size causes significant performance degradation due to increased register usage.}

\subsubsection{{\bf Q-B2:} The effect of forgettable hash table management in throughput}
\begin{figure}[t]
    \centering
    \includegraphics[width=\linewidth]{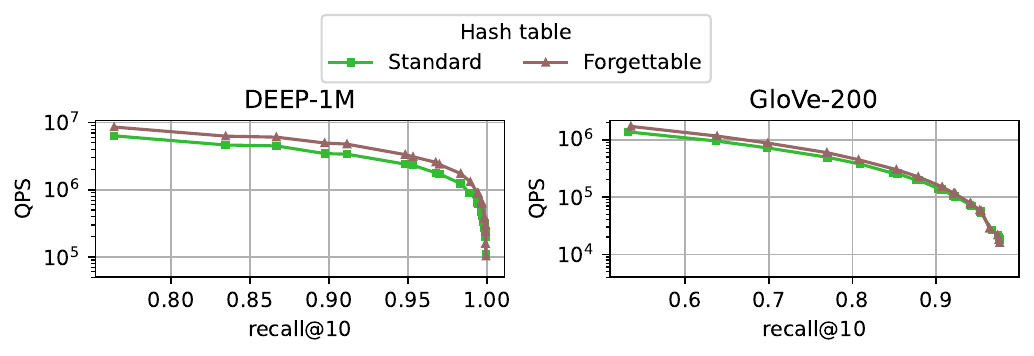}
    \caption{Search performance comparison between two hash table management methods: standard and forgettable.}
    \label{fig:hash}
\end{figure}
When we place the hash table in shared memory, we use the \textit{forgettable hash table management}, a small-size hash table that is reset periodically, instead of the standard hash table in device memory, as mentioned in Sec. \ref{sec:hash}.
Then, \textit{how much faster is the forgettable hash on shared memory than the standard one on the device memory}, and \textit{how much recall is reduced by the periodic reset?}
We compare the search performance between the forgettable hash in shared memory and the standard hash in device memory when using them in single-CTA in Fig. \ref{fig:hash}.
In this experiment, we reset the hash table for every iteration in the forgettable hash.
In both datasets, DEEP-1M and GloVe, we have confirmed the forgettable hash achieves compatible or higher search throughput compared to the standard hash.
The throughput gain observed in GloVe is slightly smaller than in DEEP-1M.
This discrepancy can be attributed to the fact that in GloVe, the overhead of hash table operations becomes relatively smaller when dealing with larger dimension dataset vectors, as the primary computational load shifted towards distance calculations.

\subsubsection{{\bf Q-B3:} Search performance comparison between single- and multi-CTA implementations}
\begin{figure}[t]
    \centering
    \includegraphics[width=\linewidth]{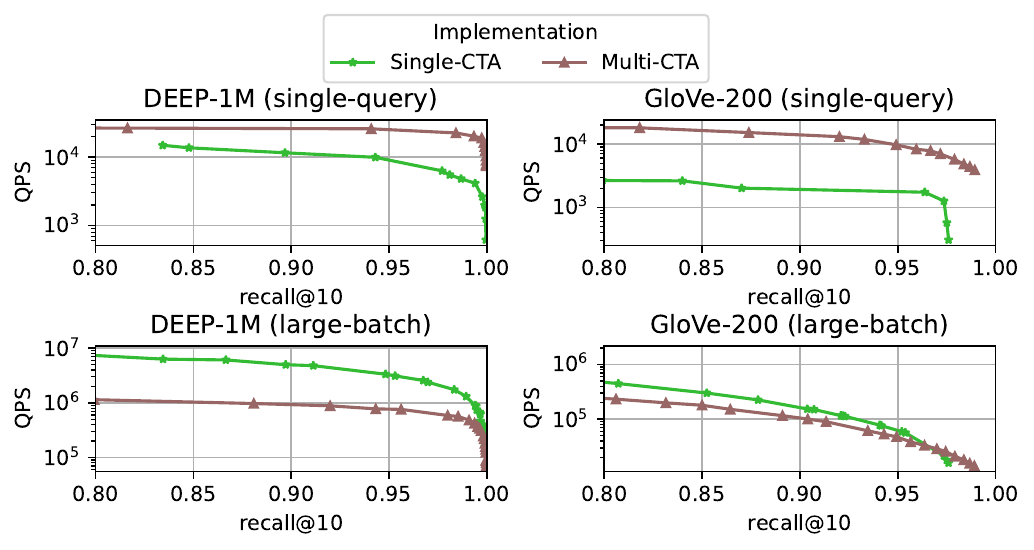}
    \caption{Search performance comparison between single-CTA and multi-CTA implementations for single-query (top) and large-batch query (bottom). The batch size in large-batch query search is 10K.}
    \label{fig:single-multi-cta}
\end{figure}
We have measured the search performance of the single- and multi-CTA implementations for the DEEP-1M and GloVe datasets, as shown in Fig. \ref{fig:single-multi-cta}.
In the context of a single query, the multi-CTA approach outperforms the single-CTA approach for both the DEEP-1M and GloVe datasets.
However, in a large-batch query, we observe divergent outcomes.
For the DEEP-1M dataset, the single-CTA method demonstrates superior search performance.
On the other hand, in the case of GloVe, if a higher recall is required, the multi-CTA method achieves better results.
This discrepancy can be attributed to the nature of the GloVe dataset, which is considered to be ``harder'' than DEEP-1M \cite{li_approximate_2020}.
Achieving higher recall on the GloVe dataset necessitates searching through more nodes, in other words, increasing internal top-$M$, which is a requirement that the multi-CTA approach fulfills effectively.

\section{Experiments}
We compare CAGRA with the following graph-based ANNS implementations:
\begin{enumerate}
    \item {\bf GGNN} \cite{groh_ggnn_2022}: One of the current state-of-the-art GPU implementation candidates.
    \item {\bf GANNS} \cite{yu_gpu-accelerated_2022}: One of the current state-of-the-art GPU implementation candidates.
    \item {\bf HNSW} \cite{malkov_efficient_2018}: Well-known state-of-the-art implementation and proximity graph for CPU.
    \item {\bf NSSG} \cite{fu_high_2022}: An implementation with search and graph construction processes similar to CAGRA.
    NSSG also starts the search process with random sampling.
\end{enumerate}
While the out-degree of the CAGRA graph is fixed, it is not fixed for the other four implementations.
Therefore, we align the average out-degree for each dataset to make the comparison between them as fair as possible.

In the experiments, we answer the following questions:
\begin{enumerate}[label={\bf Q-C\arabic*},leftmargin=3em]
    \item How fast is the CAGRA graph construction?
    \item How comparable is the search quality of the CAGRA graph?
    \item How much better is the CAGRA search performance in batch processing?
    \item How much better is the CAGRA search performance in online processing?
    \item Does CAGRA support large datasets?
\end{enumerate}

\subsection{{\bf Q-C1:} Graph construction time}
\begin{figure}[t]
    \centering
    \includegraphics[width=\linewidth]{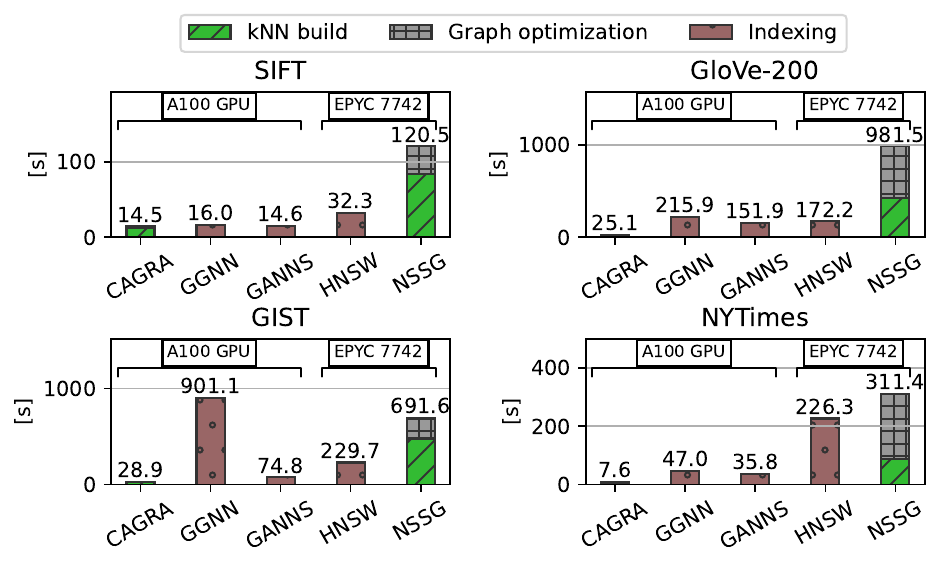}
    \caption{Graph construction time comparison among CAGRA and other graph-based ANNS implementations.}
    \label{fig:construction-time}
\end{figure}
The result of the graph construction time is shown in Fig. \ref{fig:construction-time}.
We measured the entire graph construction time, including memory allocation, dataset file load, and data movement.
NSSG first builds a $k$-NN graph explicitly and then optimizes it similarly to CAGRA, whereas GGNN, GANNS, and HNSW do not.
Therefore, in the case of CAGRA and NSSG, we show the breakdown of the initial $k$-NN graph build and its optimization time, while the others are only the entire construction time.
CAGRA is compatible with or faster than the other CPU and GPU implementations.
In comparing implementations for GPU, CAGRA is $1.1 \textendash{} 31\times$ faster than GGNN, and $1.0\textendash{} 6.1\times$ than GANNS. And it is $2.2\textendash{} 27\times$ faster than HNSW.

\subsection{{\bf Q-C2:} Graph search quality}
\begin{figure}[t]
    \centering
    \includegraphics[width=\linewidth]{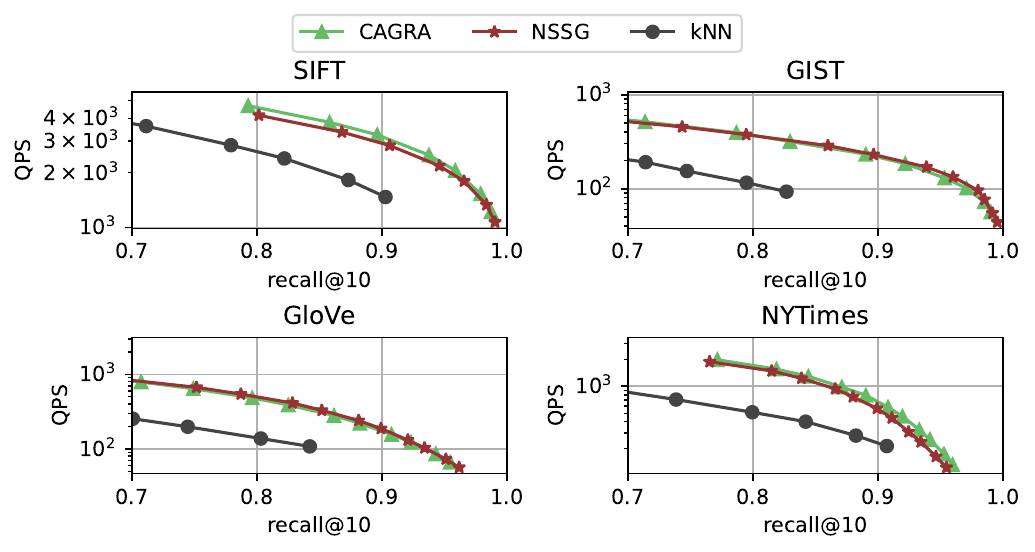}
    \caption{The search performance comparison between graphs created by CAGRA and NSSG using NSSG single-threaded search implementation.}
    \label{fig:NSSG-comparison}
\end{figure}
To evaluate the search quality of the graph, we compare the search performance of the CAGRA graph to the NSSG graph using the NSSG search implementation in both graphs.
To do so, we load the CAGRA graph into NSSG and use NSSG search to find nearest neighbors. 
\revised{Using the same search implementation with different graphs allows us to directly compare the quality of the graphs. Fig. \ref{fig:NSSG-comparison} demonstrates that while a kNN graph results in low search accuracy, the CAGRA and NSG graphs show comparable performance.}
Although many graphs have high performance, we use NSSG since its search implementation is similar to CAGRA, including that the search process starts from the random sampling, and it has better or compatible search performance compared to most of them \cite{wang_comprehensive_2021}.
If we were to use another search implementation unsuitable for CAGRA, for instance, HNSW or NSG, the search performance would be disadvantageous to CAGRA.
In this evaluation, we first build an NSSG graph and calculate the average out-degree of the graph.
Then, we build a CAGRA graph for the dataset setting the out-degree as the largest value less than or equal to the average out-degree in multiples of $16$.
The comparison of search performance is shown in Fig. \ref{fig:NSSG-comparison}, and the results indicate that the search performance of the CAGRA graph is almost at the same level as that of the NSSG graph across all four datasets.
In summary, the evaluation demonstrates that the CAGRA graph achieves a search performance comparable to the NSSG graph, which is one of the highest-performance graphs for ANNS.

\subsection{{\bf Q-C3:} Recall and throughput in large-batch query search}

\begin{figure}[t]
    \centering
    \includegraphics[width=1.\linewidth]{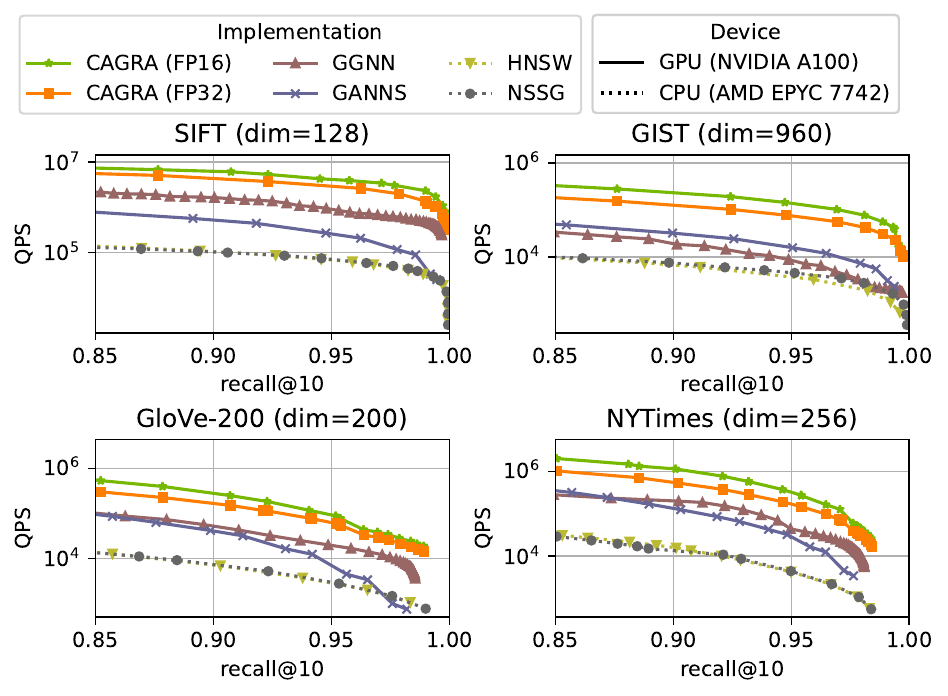}
    \caption{
    Large-batch search performance comparison among CAGRA and other graph-based ANNS implementations (batch size=10K).
    CAGRA (FP32) indicates that the dataset is stored in FP32, while CAGRA (FP16) is converted to FP16.
    }
    \label{fig:multi-batch-recall-throughput}
\end{figure}
In the batch processing use of ANNS, large-batch search performance is crucial.
This use case is suitable for GPU since it is easy to extract parallelism, and this makes the best use of the single-CTA implementation in CAGRA.
Then, \textit{How fast is CAGRA compared to other ANNS GPU implementations and the state-of-the-art CPU implementation in large-batch search?}
We have compared the recall and throughput among CAGRA and the other methods, as shown in Fig. \ref{fig:multi-batch-recall-throughput}.
Since the search implementation of NSSG is not multi-threaded, and using it would not be a fair comparison, we measured the performance of NSSG using the search implementation for the bottom layer of the HNSW graph.
In the performance measurement of HNSW and NSSG, we have tried multiple OpenMP thread counts, up to 64, and plotted the fastest of them.
The results show that the performance of CAGRA is higher than the other ANNS methods on both CPU and GPU.
In the 90\% to 95\% recall range, our method is $33\textendash{} 77\times$ faster than HNSW and is $3.8\textendash{} 8.8\times$ faster than the other GPU implementations.
Since the memory bandwidth of the device limits the throughput of CAGRA, we can gain higher throughput using half-precision (FP16) in dataset vector data type.
We demonstrate that half-precision does not degrade the quality of results while still benefitting from higher throughput.

\subsection{{\bf Q-C4:} Recall and throughput in single-query search}
\begin{figure}[t]
    \centering
    \includegraphics[width=\linewidth]{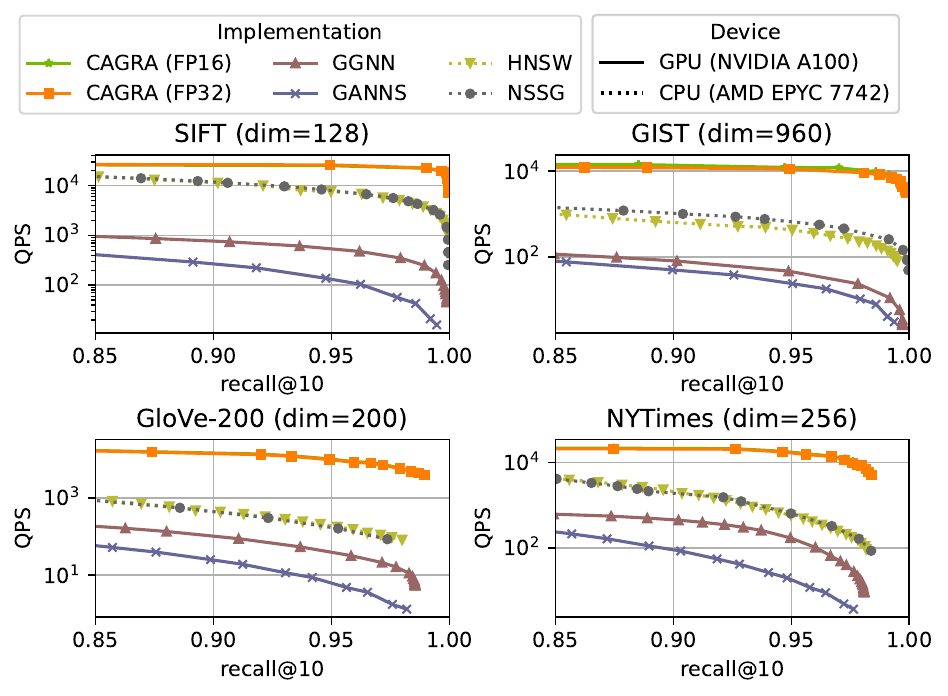}
    \caption{
    Single-query search performance comparison among CAGRA and other graph-based ANNS implementations.
    CAGRA (FP32) indicates that the dataset is stored in single-precision FP32, while CAGRA (FP16) is converted to half-precision FP16.
    }
    \label{fig:single-batch-recall-throughput}
\end{figure}
In the online processing use of ANNS, single-query performance is crucial.
In this use case, an implementation for multi-batch processing on GPU is typically unsuitable since it can not efficiently utilize the GPU resources.
In CAGRA, we propose the multi-CTA implementation to address this inefficiency.
Then, \textit{How fast is CAGRA compared to the other fast ANNS implementations for CPU in single-query?}
\revised{We have compared the recall and throughput of CAGRA to the other methods}, as shown in Fig. \ref{fig:single-batch-recall-throughput}.
Our results show that CAGRA has a $3.4\textendash{} 53\times$ higher search performance than HNSW at 95\% recall.
Since the GGNN and GANNS methods are optimized for large-batch queries, \revised{their single-query throughputs are much slower than even HNSW and NSSG on CPU}.
While the performance of CAGRA (FP16) and CAGRA (FP32) in SIFT, GloVe, and NYTimes are very similar, CAGRA (FP16) is slightly better in GIST.
This discrepancy is from the larger dimensionality of GIST compared to the other datasets, which require more memory bandwidth to load the dataset.

\subsection{{\bf Q-C5:} Large size dataset}
\begin{figure}[t]
    \centering
    \includegraphics[width=\linewidth]{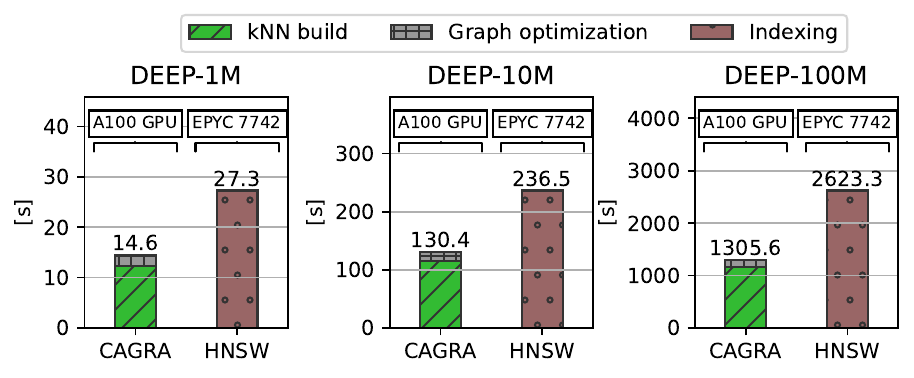}
    \caption{
    Graph construction time comparison between CAGRA and HNSW for DEEP-1M, 10M, and 100M datasets.
    }
    \label{fig:deep-large-construction}
\end{figure}
\begin{figure}[t]
    \centering
    \includegraphics[width=\linewidth]{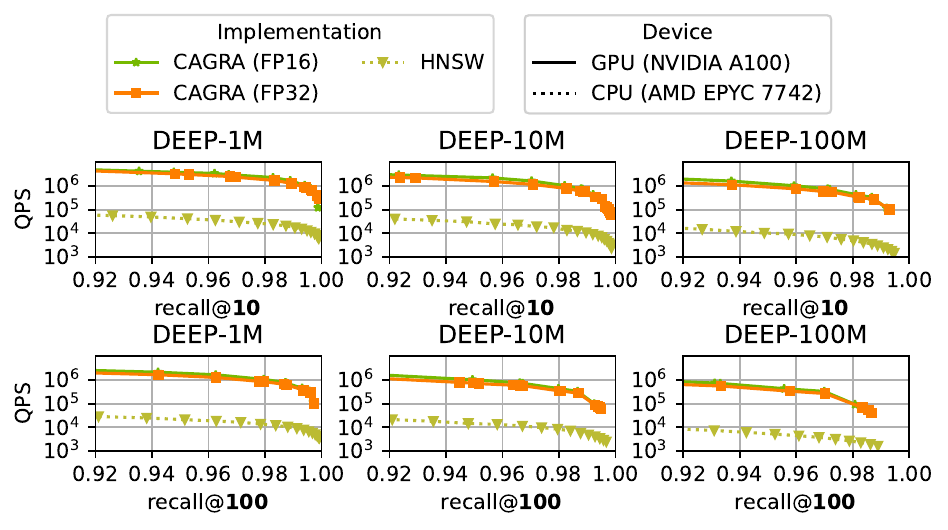}
    \caption{
    Search performance comparison between CAGRA and HNSW for DEEP-1M, 10M, and 100M datasets in recall@10 (top) and recall@100 (bottom).
    The batch size is 10K.
    }
    \label{fig:deep-large}
\end{figure}
In recent years, the query operations for larger and larger datasets are attracting attention \cite{simhadri_results_2022}.
So, \textit{does CAGRA support large datasets?}, and \textit{how do larger datasets affect CAGRA's search performance?}
We measured the graph construction and search performance of CAGRA in the DEEP-1M, 10M, and 100M datasets, as shown in Fig. \ref{fig:deep-large-construction} and Fig. \ref{fig:deep-large}.
The graph construction time increases proportionally with the dataset size.
During the search performance comparisons, we observe that as the dataset size grows, CAGRA's recall declines slightly but follows a similar trend to HNSW and the degradation in both recall and throughput is not significant.
Based on our findings, we believe that CAGRA remains capable of handling larger datasets while maintaining this trend unless the dataset exceeds the device memory capacity.
In these cases, a multi-GPU sharding technique \cite{doshi_lanns_2020} discussed in Sec. \ref{sec:multi_cta_impl} \revised{and data compression schemes, such as product quantization, are some of the ways to address the memory capacity problem, though further performance investigation is required}.

\section{Conclusion}
In this paper, we proposed a fast graph-based ANNS method called CAGRA\revised{, which is designed to perform well on NVIDIA GPUs by harnessing their increased computing capacity and superior memory bandwidth}.
CAGRA performs a heuristic optimization to the initial $k$-NN graph to improve the reachability from each node to other nodes with a highly parallel computation-friendly algorithm. 
CAGRA has better search performance than other state-of-the-art graph-based ANNS implementations on both CPU and GPU for both large-batch and single queries.
CAGRA is available in the open-source NVIDIA RAPIDS RAFT library, which can be found on GitHub (\url{https://github.com/rapidsai/raft}).

\bibliographystyle{plain}
\bibliography{ref}

\end{document}